\documentclass[nofootinbib,aps,preprintnumbers,unsortedaddress,prl,twocolumn]{revtex4}
\pdfoutput=1
\usepackage{graphicx}
\usepackage{textcomp}
\usepackage{bm}
\usepackage{times}
\usepackage{color}
\usepackage{graphics}
\usepackage{hyperref}
\usepackage{setspace}
\usepackage{paralist}
\usepackage{bbm}
\usepackage{color}
\usepackage{slashed}
\usepackage{comment}

\usepackage{rotating}
\usepackage{framed}

\usepackage{placeins}

\usepackage{amsmath} 
\usepackage{amsthm} 
\usepackage{amssymb}	
 
\let\baraccent=\= 
\renewcommand{\=}[1]{\stackrel{#1}{=}} 

\theoremstyle{definition}

\theoremstyle{remark}

\usepackage{bbm}

\hypersetup{
    pdfnewwindow=true,      
    colorlinks=true,       
    linkcolor=black,          
    citecolor=blue,        
    filecolor=blue,      
    urlcolor=blue           
}

\makeatletter
\newcommand\xleftrightarrow[2][]{%
  \ext@arrow 9999{\longleftrightarrowfill@}{#1}{#2}}
\newcommand\longleftrightarrowfill@{%
  \arrowfill@\leftarrow\relbar\rightarrow}
\makeatother

\begin{document}
\title{\Large {\it{\bf{Dark Matter and The Seesaw Scale}}}}
\author{Pavel Fileviez P\'erez$^{1}$, Clara Murgui$^{2}$}
\affiliation{$^{1}$Physics Department and Center for Education and Research in Cosmology and Astrophysics (CERCA), Case Western Reserve University, Rockefeller Bldg. 2076 Adelbert Rd. Cleveland, OH 44106, USA \\
$^{2}$Departamento de F\'isica Te\'orica, IFIC, Universitat de Valencia-CSIC, 
E-46071, Valencia, Spain}
\begin{abstract}
We discuss the possibility to find an upper bound on the seesaw scale using 
the cosmological bound on the cold dark matter relic density. We investigate a simple relation between the 
origin of neutrino masses and the properties of a dark matter candidate in a simple theory where the 
new symmetry breaking scale defines the seesaw scale. Imposing the cosmological bounds, we find 
an upper bound of order multi-TeV on the lepton number violation scale. We investigate the predictions 
for direct and indirect detection dark matter experiments, and the possible signatures at the Large Hadron Collider.
\end{abstract}
\maketitle 
%
\section{Introduction}
The origin of neutrino masses and the nature of the cold dark matter in the Universe are two of the most exciting open problems in particle physics and cosmology.
We know today about several mechanisms to generate neutrino masses, see for example Ref.~\cite{FileviezPerez:2009ym}, but the so-called seesaw 
mechanism~\cite{TypeI}  is considered the most appealing and simple mechanism for Majorana neutrino masses. Unfortunately, we only know that 
the upper bound on the seesaw scale is about $10^{14}$ GeV, which is an energy scale very far from any future collider experiment. 
Therefore, it is not clear we could test the mechanism behind neutrino mass. There are also many possible candidates 
to describe the cold dark matter in the Universe, see for example Ref.~\cite{Bertone:2004pz}. The weakly interacting massive particles (WIMPs) have been 
popular dark matter candidates in the last decade but the recent experimental results tell us that maybe one should think about other possibilities.
However, it is fair to say that the idea of describing the dark matter with WIMPs is so appealing that it is better to understand and revise all constraints 
and the different models before abandoning this idea. 

The discovery of lepton number violating signatures in low energy experiments or at colliders will be striking signals for new physics beyond the Standard Model.
In low energy experiments we could discover neutrinoless double beta decay, for a review see Ref.~\cite{Elliott:2002xe}, and at colliders different signatures 
with same-sign leptons could be seen~\cite{Cai:2017mow}. These discoveries will be crucial to establish the origin of neutrino masses.

We understand the origin of charged fermion masses in the Standard Model through the spontaneous breaking of the electroweak symmetry. In the same 
way, we could understand the origin of the seesaw scale if $B-L$  is a local symmetry spontaneously broken through the Higgs mechanism. Unfortunately, 
as in canonical seesaw, the upper bound on the $B-L$ is typically very large, $M_{B-L} \lesssim 10^{14}$ GeV.  There are two known ways to establish a much 
smaller upper bound on the $B-L$ breaking scale: a) In the context of the minimal supersymmetric $U(1)_{B-L}$ theory~\cite{Barger:2008wn} the $B-L$ 
breaking scale is defined by the supersymmetry breaking scale. Then, if low energy supersymmetry is realized at the multi-TeV scale, we could 
discover lepton number violating signatures at colliders. b) The second possibility is to use the cosmological bounds on the dark matter relic 
density to impose an upper bound on the $B-L$ breaking scale in the case where the dark matter is charged under the same gauge symmetry.
\begin{figure}[t]
\includegraphics[width=0.8\linewidth]{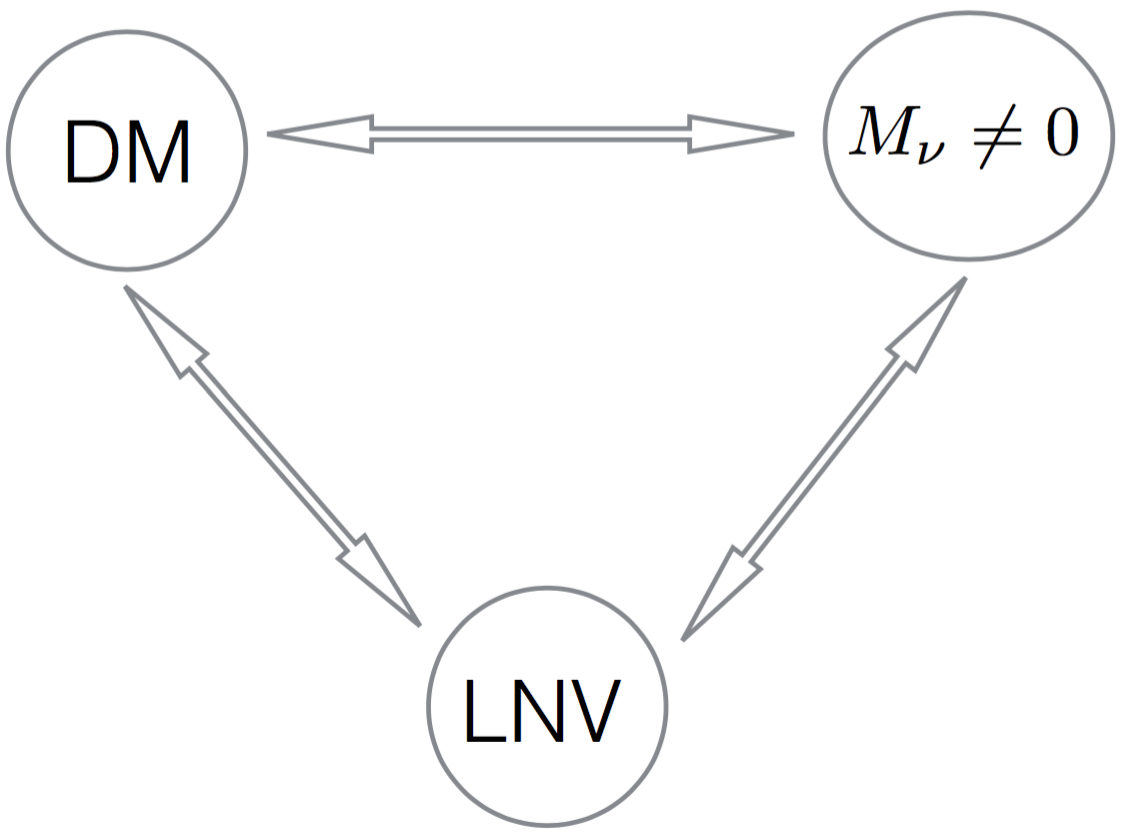}
\caption{Correlation between the origin of neutrino masses, properties of cold dark matter candidate and lepton number violating signatures.}
\end{figure}

In this article, we focus on the second possibility mentioned above in order to find an upper bound on the $B-L$ seesaw scale. 
In this theory, the dark matter candidate is a vector-like fermion which is a SM singlet but charged under the $B-L$ gauge symmetry. 
We find that, using the constraints on the cold dark matter relic density, the upper bound on the $B-L$ is in the multi-TeV region. 
Therefore, one can expect exotic signatures at colliders with same-sign multi-leptons and displaced vertices. {\textit{This connection between 
the cosmological dark matter bounds and exotic signatures at colliders is very unique and one could hope to test the origin of neutrino masses 
at colliders}}. See Fig.~1 for a simple way to illustrate this correlation.

\section{ Neutrino Masses and The $B-L$ Scale}
The simplest gauge theory where one can understand dynamically the origin of neutrino masses is based on the $B-L$ gauge symmetry.
In this context, we add three copies of right-handed neutrinos to define an anomaly free theory, and one can easily implement the 
seesaw mechanism~\cite{TypeI} for Majorana neutrino masses. The relevant Lagrangian for the generation of neutrino mass is given by
\begin{equation}
-{\cal{L}}_{\rm{Seesaw}} =  Y_\nu \ \bar{\ell}_L i \sigma_2 H^* \nu_R \ + \ \lambda_R \ \nu_R^T C \nu_R S_{BL} + \rm{h.c.},
\end{equation} 
where $\nu_R \sim (1,1,0,-1)$ are the right-handed neutrinos, $H \sim (1,2,1/2,0)$ is the Standard Model Higgs, and 
$S_{BL} \sim (1,1,0,2)$ is the new Higgs responsible for the spontaneous breaking of $B-L$. Using the above interactions 
one can generate masses for the Standard Model neutrinos through the well-known Type I seesaw mechanism~\cite{TypeI}, which are given by
\begin{eqnarray}
M_\nu^I &=&  m_D^T M_R^{-1} m_D,
\end{eqnarray}
where $m_D=Y_\nu v_0/\sqrt{2}$. Here $v_0/\sqrt{2}$ is the vacuum expectation value of the Standard Model Higgs. 
We note that, in this case, the masses of the right-handed neutrinos, $M_R = \sqrt{2} \lambda_R v_{BL}$, are defined 
by the $B-L$ breaking scale. The seesaw scale, in general, is unknown; the only thing we know is that the upper bound should be the canonical seesaw scale $10^{14}$ GeV. It is important to mention that if $M_R$ is at the TeV scale one can generate masses for Standard Model neutrinos in agreement with the experiments if $m_D < 10^{-3}$ GeV, and one can produce the right-handed neutrinos at the LHC through the $B-L$ neutral gauge boson, $pp \to Z_{BL}^* \to N_i N_i$~\cite{Basso:2008iv,Huitu:2008gf,AguilarSaavedra:2009ik,Accomando:2017qcs,Perez:2009mu}, giving rise to striking lepton number violating signatures with same-sign leptons and multijets.

Now, since the observation of lepton number violation is crucial to learn about the origin of neutrino masses, it is important 
to understand the possibility to find an upper bound on the $B-L$ breaking scale which is much smaller than the canonical seesaw scale. 
Then, we could hope to test the origin of neutrino masses at current or future experiments. We know about two different class of theories where it is possible to find an upper bound on the $B-L$ breaking scale:

\begin{itemize}

\item In Ref.~\cite{Barger:2008wn}, one of us (P.F.P.) and collaborators pointed out that in the minimal 
supersymmetric $B-L$ model the gauge symmetry must be broken by the vacuum expectation value 
of the `right-handed' sneutrinos. Then one predicts that $R-$parity must be spontaneously broken, and 
one expects the existence of lepton number violation. In this context, the $R-$parity and 
lepton number violation scales are defined by the supersymmetry breaking scale. 
Then, if one has low energy supersymmetry at the multi-TeV scale, and this theory 
is true, one should discover lepton number violation at current or future colliders.
For detailed studies see Refs.~\cite{FileviezPerez:2012mj,FileviezPerez:2008sx,Perez:2013kla,Barger:2010iv}.

\item The second possibility is discussed in details in this article. We will show that if one has a 
fermionic cold dark matter candidate which is charged under the $B-L$ gauge symmetry, 
it is possible to find an upper bound on the $B-L$ breaking scale in the multi-TeV region 
using the cosmological bounds on the dark matter relic density. Therefore, this theory provides 
a simple scenario which motivates the search for lepton number violation at colliders. 

\end{itemize}

These two scenarios provide two major examples of theories where one could expect the discovery of lepton number violating processes at the multi-TeV scale.
We will focus on the second example and investigate the impact of all dark matter bounds.

\section{Dark Matter and the $B-L$ Scale}
One can write a very simple model to generate Majorana neutrino masses and to explain the presence of cold dark matter in the Universe based 
on the spontaneous breaking of the $U(1)_{B-L}$ gauge symmetry. The relevant Lagrangian for our discussion is given by
\begin{eqnarray}
{\cal{L}}_{\nu}^{DM} &\supset& - \frac{1}{4} F^{BL}_{\mu \nu} F^{BL}_{\alpha \beta}  g^{\alpha \mu} g^{\beta \nu} +  i \overline{\chi}_L \gamma^\mu D_\mu \chi_L +  i \overline{\chi}_R \gamma^\mu D_\mu \chi_R  \nonumber \\
&+ & (D_\mu S_{BL})^\dagger (D^\mu S_{BL})  - ( Y_\nu \ \bar{\ell}_L i \sigma_2 H^* \nu_R \nonumber \\
 & + & \ \lambda_R \ \nu_R^T C \nu_R S_{BL}  +  M_\chi \bar{\chi}_L \chi_R + \rm{h.c.} ),
\end{eqnarray} 
where $F^{BL}_{\mu \nu} = \partial_\mu Z_{BL\nu} - \partial_\nu Z_{BL\mu}$ defines the kinetic term for the $B-L$ gauge boson $Z_{BL}$. Since $\chi_L \sim (1,1,0,n)$ 
and $\chi_R \sim (1,1,0,n)$, the covariant derivates are defined by $D^\mu \chi_L= \partial^\mu \chi_L + i g_{BL} n Z_{BL}^\mu \chi_L$ and 
$D^\mu \chi_R= \partial^\mu \chi_R + i g_{BL} n  Z_{BL}^{\mu} \chi_R$. Here, $|n|\neq 1,3$  in order to avoid the decay of $\chi=\chi_L + \chi_R$, which must be stable, and can be a good cold dark matter candidate. In the case the proposed theory is sensitive to UV physics, we note that mixing among neutrinos and the dark matter candidate could be originated by non-renormalizable operators only for a choice of $n$ odd, whereas even and fractionally charges would be safe. The kinetic mixing between the $B-L$ gauge boson and the hypercharge gauge boson is neglected for simplicity.
A similar model has been partially investigated before in Ref.~\cite{Duerr:2015wfa}, where the main emphasis was the study of the gamma lines from dark matter annihilation.
Our main goal here is to investigate in detail the connection between the cosmological bounds and the lepton number violation scale, and understand the implications for the search for lepton number violation at colliders.

\textit{Higgs Sector:}
The Higgs sector of this theory is composed of the SM Higgs $H \sim (1,2,1/2,0)$ and $S_{BL} \sim (1,1,0,2)$ and
the scalar potential is given by 
\begin{eqnarray}
V(H,S_{BL}) &=& - \mu_H^2 H^\dagger H + \lambda_H \left( H^\dagger H \right)^2 - \mu_{BL}^2 S_{BL}^\dagger S_{BL} \nonumber \\
&+ & \lambda_{BL}  \left( S_{BL}^\dagger S_{BL} \right)^2 + \lambda_{HBL} \left( H^\dagger H \right) \left( S_{BL}^\dagger S_{BL} \right), \nonumber \\
\end{eqnarray}
where
\begin{eqnarray}
H^T &=& \left( \begin{array}{cc}
H^+, \ \displaystyle  \frac{1}{\sqrt{2}} \left( v_0 + h + i A \right) 
\end{array} \right) \  {\rm{and}} \nonumber \\  
S_{BL} &=& \frac{1}{\sqrt{2}} \left( v_{BL} + h_{BL} + i A_{BL} \right). 
\end{eqnarray}
In this case the physical states are:
\begin{eqnarray}
h_1 &=& h \cos \theta_{BL} + h_{BL} \sin \theta_{BL}, \\
h_2 &=& - h \sin \theta_{BL} + h_{BL} \cos \theta_{BL},
\end{eqnarray}
where
\begin{equation}
\tan 2 \theta_{BL} = \frac{ \lambda_{HBL} v_0 v_{BL}}{\lambda_H v_0^2 - \lambda_{BL} v_{BL}^2}.
\end{equation}
After symmetry breaking one finds that the mass for the $B-L$ gauge boson is given by
\begin{eqnarray}
M_{Z_{BL}} = 2 g_{BL} v_{BL}. 
\end{eqnarray}
%
The dark matter candidate $\chi=\chi_L + \chi_R$ is a Dirac fermion with mass $M_\chi$.
We focus on the case where the dark matter candidate is a Dirac fermion because in the other cases, scalar or Majorana fermion, the annihilation 
cross section through the $B-L$ gauge boson is suppressed. 
This simple dark matter model has the following relevant parameters for the dark matter study:
\begin{equation}
n, \ g_{BL}, \ M_\chi, \  M_{h_2} \ M_{Z_{BL}}, \  \text{and} \ M_{N_i} \ (i=1,2,3).
\end{equation}
As we have mentioned above the properties of the DM candidate $\chi$ are very simple since it is a vector-like 
fermion, $\chi=\chi_L + \chi_R$, and it has interactions only with the $B-L$ gauge boson. The two body annihilation channels are
\begin{eqnarray}
\bar{\chi} \chi &\to& Z_{BL}^* \to \bar{u}_i u_i, \bar{d}_i d_i, \bar{e}_i e_i, \bar{\nu}_i \nu_i, \bar{N}_i N_i, \\
\bar{\chi} \chi & \to & Z_{BL} Z_{BL}, Z_{BL} h_{1}, Z_{BL} h_2,
\end{eqnarray}
where the first one is the dominant channel when $2 M_\chi < M_{Z_{BL}}$, and the second channels are possible when $M_\chi > M_{Z_{BL}}$, $2 M_\chi > M_{Z_{BL}} + M_{h_1}$ or $2 M_\chi > M_{Z_{BL}} + M_{h_{2}}$, respectively. The  3-body annihilation channels could be important, $\bar{\chi} \chi \to Z_{BL} Z_{BL}^*$, when $M_{Z_{BL}}/2 \leq M_\chi < M_{Z_{BL}}$.
%
\begin{figure}[t]
\includegraphics[width=0.8\linewidth]{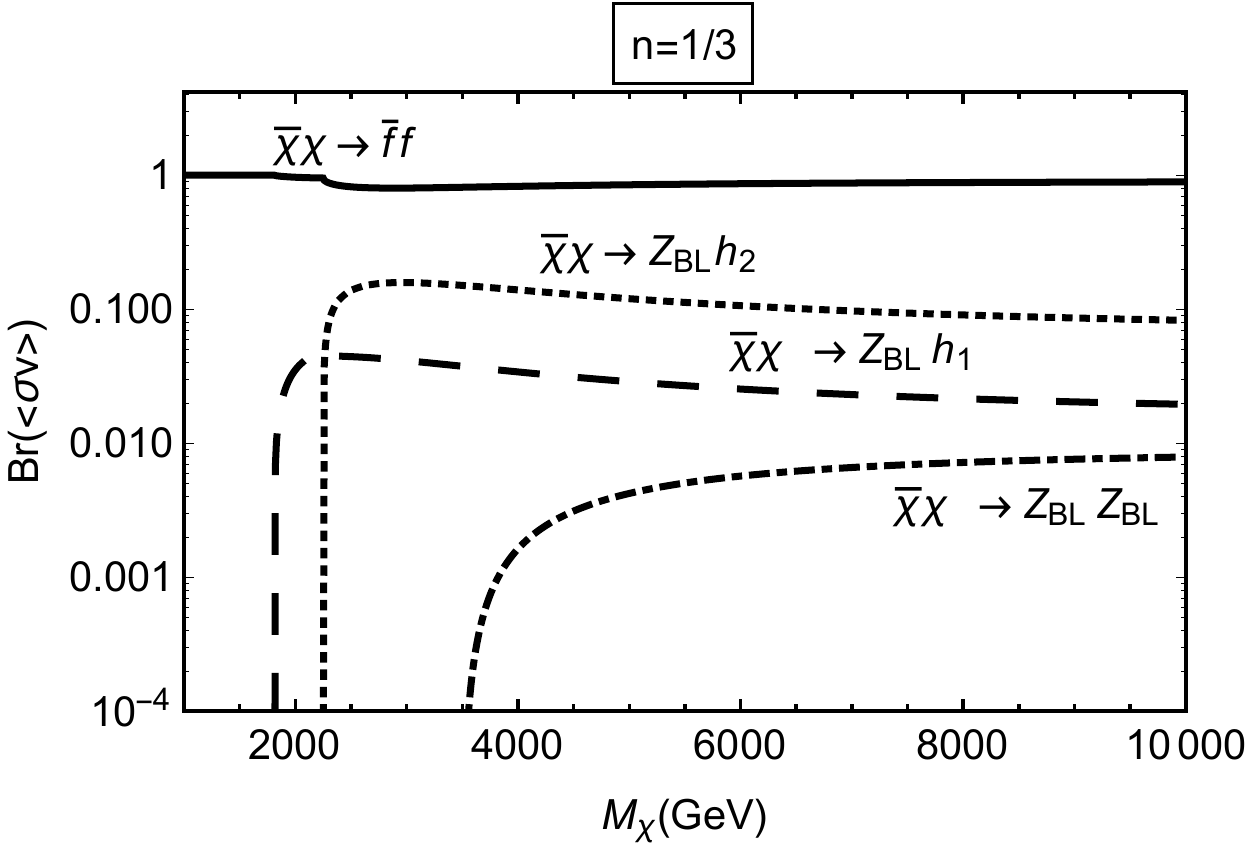}
\includegraphics[width=0.8\linewidth]{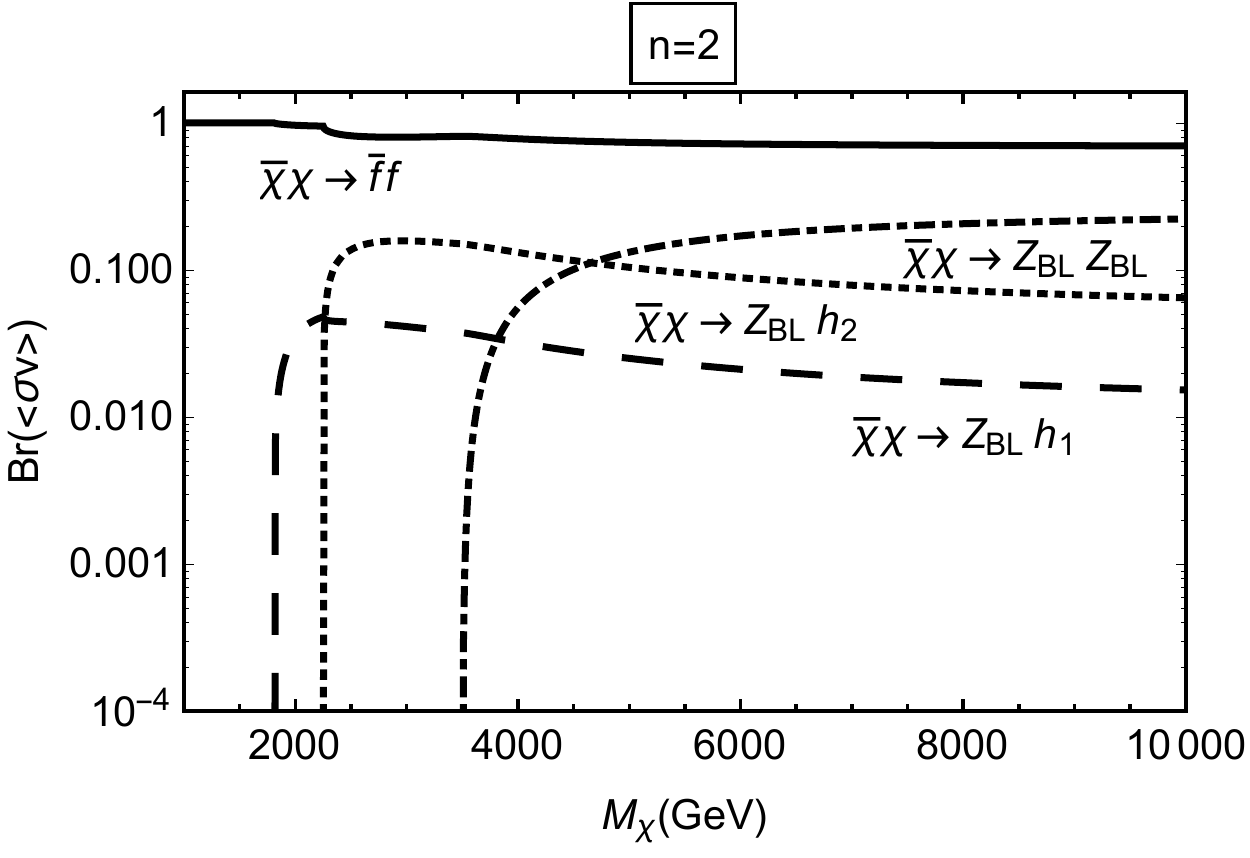}
\caption{Branching ratios of the thermal averaged cross-sections for the channels $\bar{\chi} \chi \to \bar{f}_i f_i$ (solid line) and $\bar{\chi} \chi \to Z_{BL} Z_{BL}$, $\bar{\chi} \chi \to Z_{BL} h_{2}$ and  $\bar{\chi} \chi \to Z_{BL} h_1$ (dashed lines) for different $B-L$ charges. Here we have taken $M_{Z_{BL}}=3.5$ TeV, $g_{BL}=0.5$, $\cos \theta_{BL}=0.9$ for the scalar mixing angle and $M_{h_{2}}=M_N=1$ TeV. In the upper-panel we show the results for $n=1/3$, while in the lower-panel we show the results when $n=2$.}
\end{figure}
\begin{itemize}
\item Relic Density

In order to compute the relic density, we use the analytic approximation~\cite{Gondolo:1990dk}
\begin{equation}
\label{eq:relicdensity}
\Omega_\text{DM} h^2 = \frac{1.07 \times 10^{9}{\rm{GeV}}^{-1}}{J(x_f) \ M_\text{Pl}},
\end{equation}
where $M_\text{Pl}=1.22 \times 10^{19}{\rm{GeV}}$ is the Planck scale and the function $J(x_f)$ reads as
\begin{equation}
J(x_f)=\int_{x_f}^{\infty} \frac{g_\ast^{1/2}(x) \langle \sigma v \rangle (x)}{x^2} dx.
\end{equation}
where $g_\ast$ is the total number of effective relativistic degrees of freedom at the time of freeze-out. 
The thermally averaged annihilation cross section times velocity $\langle \sigma v \rangle$ is a function of $x=M_\chi/T$, and is given by
\begin{equation}
 \langle\sigma v\rangle (x) = \frac{x}{8 M_\chi^5 K_2^2(x)} \int_{4 M_\chi^2}^\infty \sigma \times ( s - 4 M_\chi^2) \ \sqrt{s} \ K_1 \left(\frac{x \sqrt{s}}{M_\chi}\right) ds,
\end{equation}
where $K_1(x)$ and $K_2(x)$ are the modified Bessel functions.
The freeze-out parameter $x_f$ can be computed using
\begin{equation}
x_f= \ln \left( \frac{0.038 \ g \ M_\text{Pl} \ M_\chi \ \langle\sigma v\rangle (x_f) }{\sqrt{g_\ast x_f}} \right),
\end{equation}
where $g$ is the number of degrees of freedom of the dark matter particle. 
\begin{figure}[t]
\includegraphics[width=0.8\linewidth]{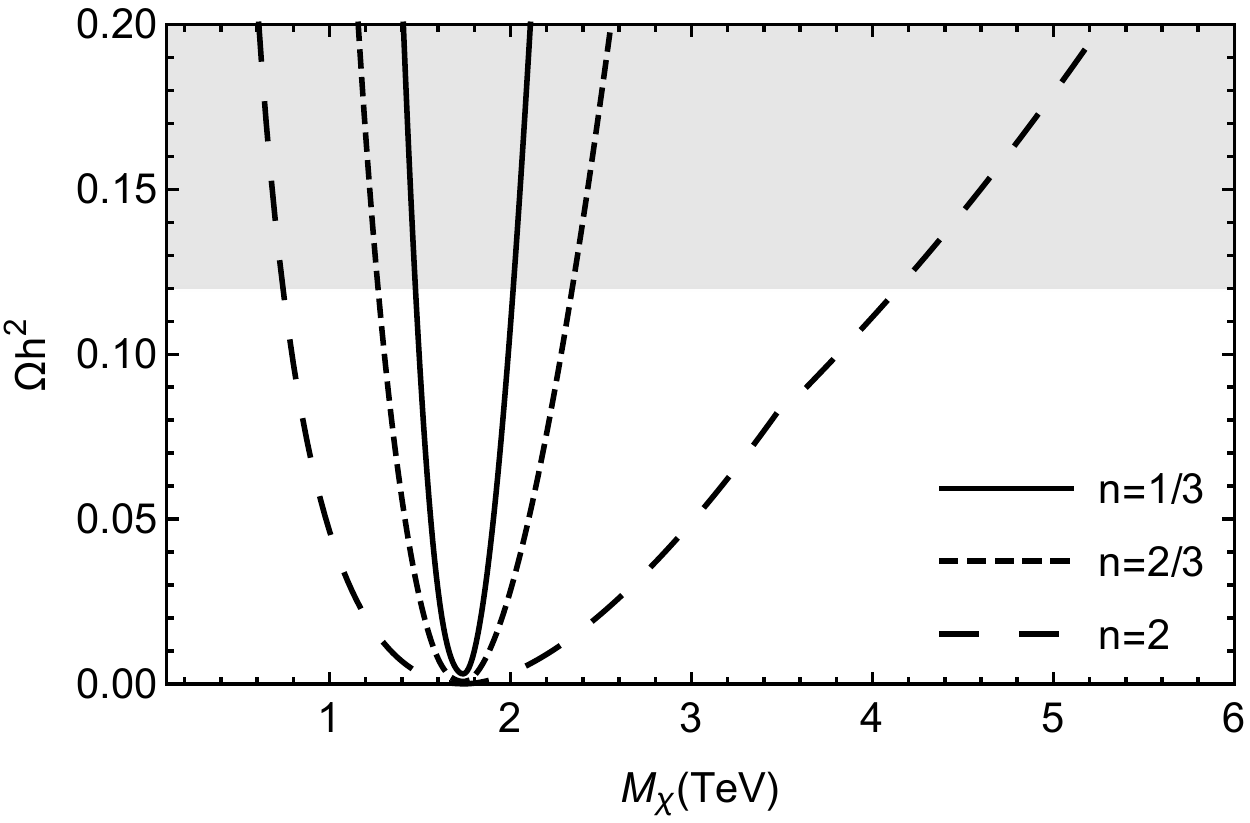}
\caption{Relic density predictions for $M_{Z_{BL}}=3.5$ TeV, $g_{BL}=0.5$ and different $B-L$ charge $n$ for the dark matter candidate. The charges $n=1/3$, $n=2/3$, and $n=2$ are represented by a solid and dashed lines, respectively. The shaded region is excluded by the bound on the relic density $\Omega h^2 \leq 0.1199 \pm 0.0027$~\cite{RelicDensity}.}
\label{channels}
\end{figure}

In Fig.~2, we show the numerical results for the branching ratios of the thermal averaged cross-sections for the channels $\bar{\chi} \chi \to \bar{f}_i f_i$ (solid line),
and $\bar{\chi} \chi \to Z_{BL} Z_{BL}$, $\bar{\chi} \chi \to Z_{BL} h_{2}$, and  $\bar{\chi} \chi \to Z_{BL} h_1$ (dashed lines), for different $B-L$ charges. 
We have used  $\cos \theta_{BL}=0.9$ for the scalar mixing angle, $M_{Z_{BL}}=3.5$ TeV, $g_{BL}=0.5$, and $M_{h_{2}}=M_N=1$ TeV for illustration. For low values of the $B-L$ charge, the annihilation channel into two fermions significantly dominates over the other channels. The annihilation into two gauge bosons $\bar{\chi} \chi \to Z_{BL} Z_{BL}$ can be important when one has large values for the dark matter $B-L$ charge. However, as we will discuss later, perturbativity bounds constrain this channel in such a way that the annihilation into fermions will always dominate over annihilation into two gauge bosons, regardless of the choice of the $B-L$ charge of the dark matter candidate. Furthermore, as we will see, it does not make sense to consider larger values of $n$ because the direct detection bounds are much stronger and one will only find consistent solutions when the gauge boson is very heavy. 

In Fig.~3, we show the predictions for the relic density when $M_{Z_{BL}}=3.5$ TeV, $g_{BL}=0.5$ and different $B-L$ charges for the dark matter candidate. 
The charges $n=1/3$, $n=2/3$, and $n=2$ are represented by a solid and dashed lines, respectively. One can see that when one has large values of the 
dark matter $B-L$ charge one can achieve the relic density in agreement with cosmology even if we are far from the resonance $M_{Z_{BL}} \approx 2M_\chi$ because 
in this case the annihilation into two gauge bosons has a larger contribution. In these studies we consider only the main annihilation channels in the numerical studies, 
and we will focus on $n=1/3$, and $n=2$ as illustrative examples.

\item Direct Detection

The elastic spin-independent nucleon--dark matter cross section is given by
\begin{equation}
\sigma_{\chi N}^\text{SI} = \frac{M_N^2 M_\chi^2}{ \pi (M_N + M_\chi)^2} \frac{g_{BL}^4}{M_{Z_{BL}}^4} n^2,
\end{equation}
where $M_N$ is the nucleon mass. We note that $\sigma_{\chi N}^\text{SI}$ is independent of the matrix elements. The cross section can be rewritten as 
\begin{equation}
 \sigma_{\chi N}^\text{SI} (\text{cm}^2) = 12.4 \times 10^{-41} \left( \frac{\mu}{1 \rm{GeV}}\right)^2 \left( \frac{1 \rm{TeV}}{r_{BL}}\right)^4 n^2 \ \text{cm}^2,
\end{equation}
where $\mu = M_N M_\chi / (M_N + M_\chi)$ is the reduced mass and $r_{BL} = M_{Z_{BL}}/g_{BL}$. In our case $M_\chi \gg M_N$, and using the collider lower bound 
$M_{Z_{BL}}/ g_{BL} >  7 \ \rm{TeV}$~\cite{Alioli:2017nzr} one finds an upper bound on the elastic spin-independent nucleon-dark matter cross section given by
\begin{equation}
\sigma_{\chi N}^\text{SI} < 4.54 \times 10^{-44}  n^2 \ \text{cm}^2,
\end{equation}
for a given value of $n$. 
\begin{figure}[t]
\includegraphics[width=0.9\linewidth]{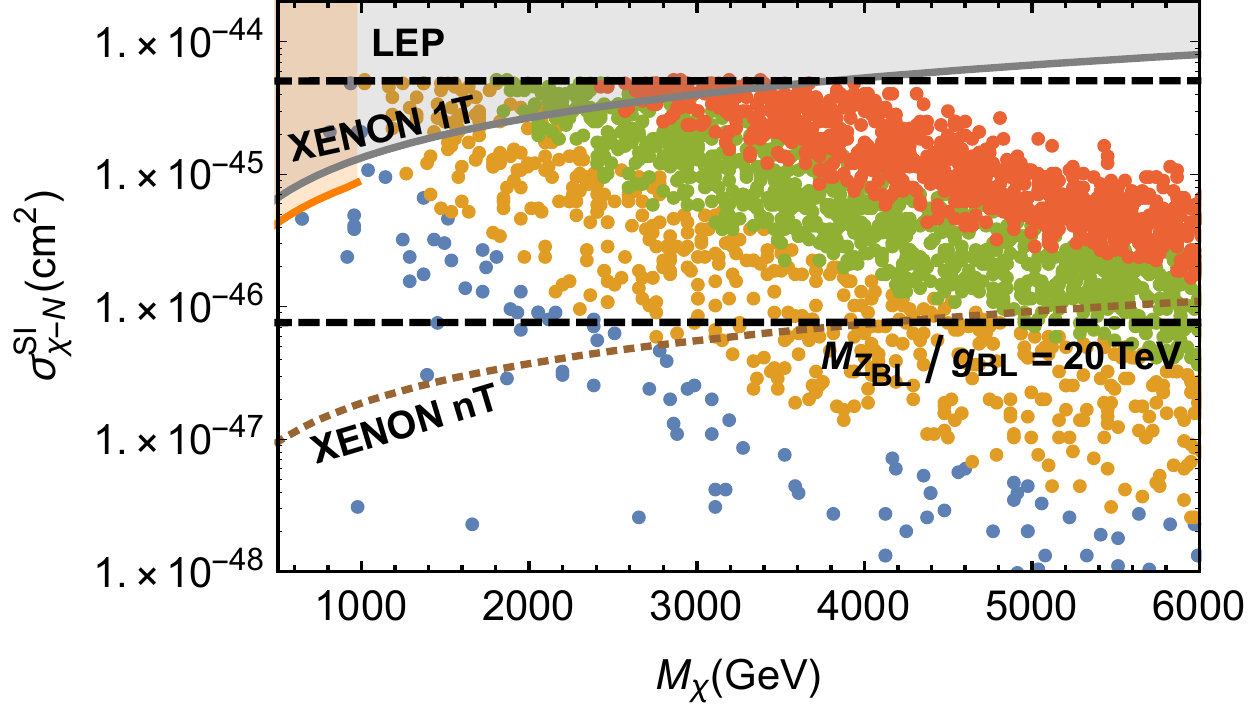}
\caption{Predictions for the direct detection spin-independent cross-section $\sigma_{\chi N}^{SI}$ for points with $n=1/3$ satisfying the relic density and LEP bounds. 
We show the Xenon-1T~\cite{Aprile:2018dbl,Aprile:2017iyp} (the last updated bounds in orange) and Xenon-nT~\cite{Aprile:2015uzo} bounds. The different colored points correspond to different values of the gauge coupling, 
$g_{BL}=0-0.25$ (blue), $g_{BL}=0.25-0.5$ (gold), $g_{BL}=0.5-0.75$ (green) and $g_{BL}=0.75-1.0$ (orange).}
\label{comparison}
\end{figure}
In Fig.~4, we show the predictions for the direct detection cross-section $\sigma_{\chi N}^{SI}$ for points with $n=1/3$ satisfying the relic density and LEP bounds. 
We show the Xenon-1T~\cite{Aprile:2018dbl,Aprile:2017iyp} and Xenon-nT~\cite{Aprile:2015uzo} bounds to understand the available region of the parameter space which 
is still in agreement with the direct detection and the expected region which could be tested in the near future. The different colored points correspond 
to the predictions when we use different values for the gauge coupling, $g_{BL}=0-0.25$ (blue), $g_{BL}=0.25-0.5$ (gold), $g_{BL}=0.5-0.75$ (green) and $g_{BL}=0.75-1.0$ (orange).
Clearly, in these scenarios the dark matter mass should be above 1 TeV to be in agreement with the direct detection bounds.

\item Indirect Detection

In this model we can have two gamma lines from dark matter annihilation, $\bar{\chi} \chi \to Z \gamma$, and $\bar{\chi} \chi \to h \gamma$. 
However, due to the fact that the cross section for the final state radiation processes $\bar{\chi} \chi \to \bar{f} f \gamma$ are much larger, one cannot identify the gamma line from the continuum spectrum.
The numerical results for these gamma lines were studied in Ref.~\cite{Duerr:2015wfa}. In Fig.~5, we show the allowed parameter space for thermal averaged dark matter annihilation cross section
into two bottom quarks (upper-panel), and two tau leptons (lower-panel), compatible with the relic density constraint. 
The gray shaded area shows the parameter space excluded by the experimental bounds from FermiLAT~\cite{Ackermann:2015zua}.
As one can see, for choices of low $n$, the allowed parameter space is compatible with these bounds. 

\begin{figure}[t]
\includegraphics[width=0.8\linewidth]{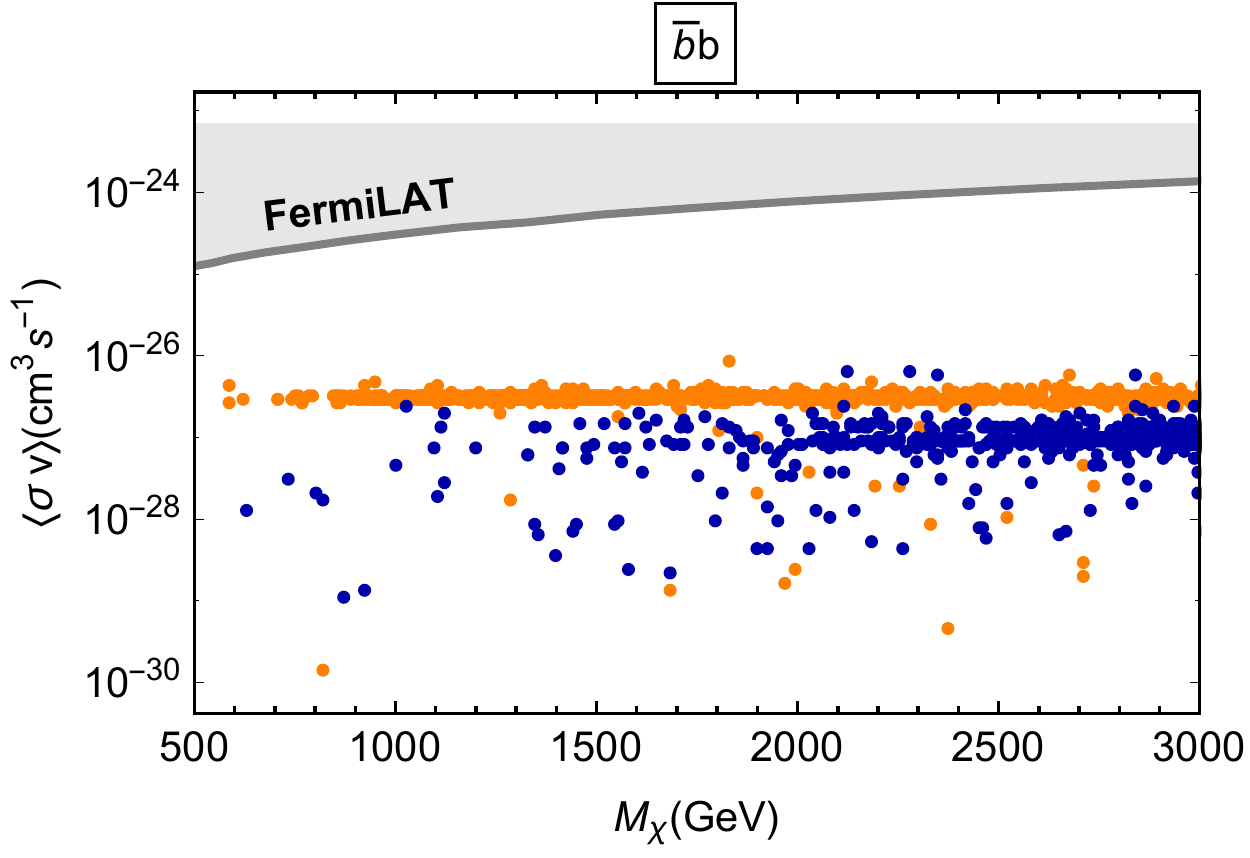}
\includegraphics[width=0.8\linewidth]{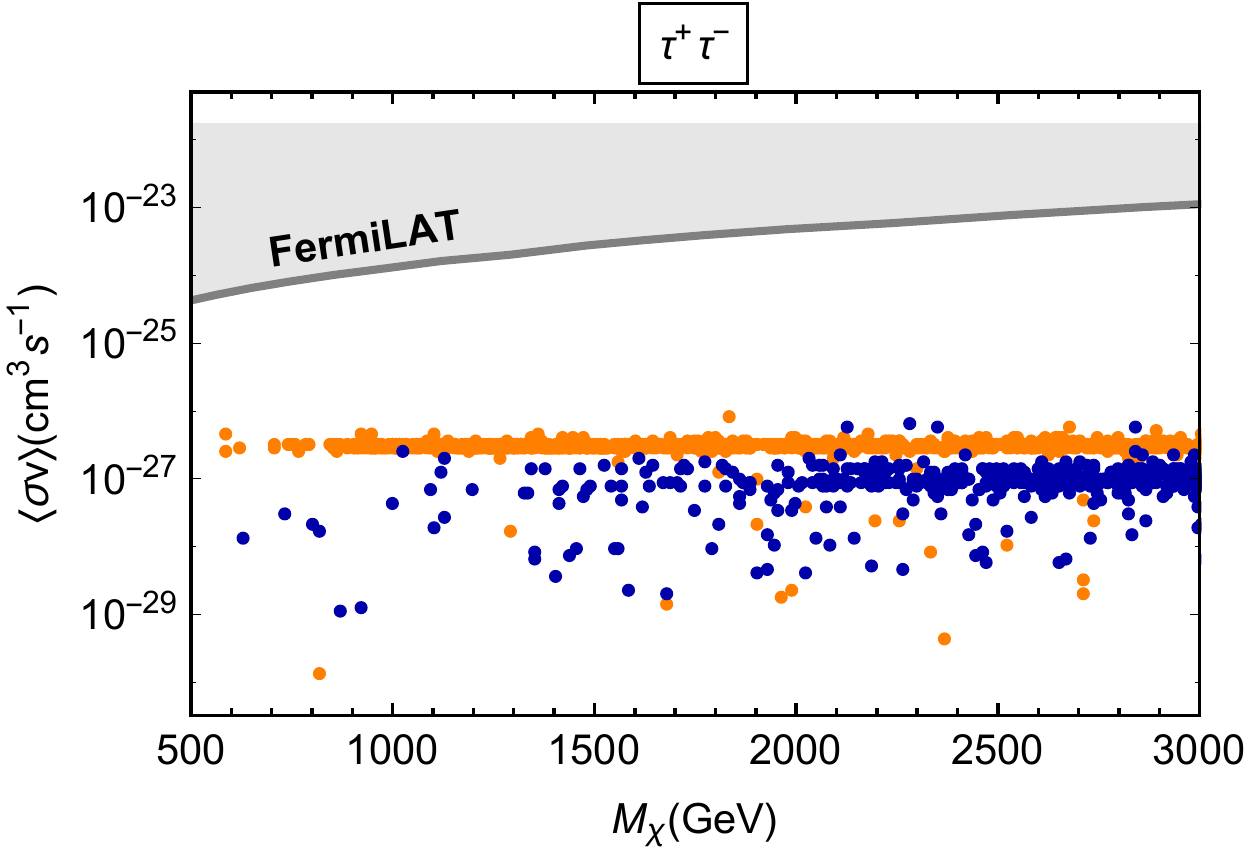}
\caption{Allowed parameter space for thermal dark matter annihilation into two bottom quarks (upper-panel) and two taus (lower-panel) compatible with the relic density constraint. The dotted-dashed lines show the predictions on the resonance for $n=1/3$ (dark blue) and $n=2$ (orange). The gray shaded area shows the parameter space excluded by the experimental bounds from FermiLAT~\cite{Ackermann:2015zua}.} 
\end{figure}

\item Upper bound on the Symmetry Breaking Scale

In Figs.~6, and~7, we show the allowed region in the $M_{Z_{BL}}-M_\chi$ plane when $\Omega_{DM} h^2 \leq 0.1199 \pm 0.027$, in agreement with the LEP bounds. 
In Fig.~6, we show the allowed solutions when the dark matter $B-L$ charge is 1/3, and the allowed region for $n=2$ is shown in Fig.~7.
As we can see, the maximum allowed value for $M_{Z_{BL}}$ is around $25$ TeV when $n=1/3$, while the upper bound on $M_{Z_{BL}}$ when $n=2$ is around 130 TeV. Clearly, each choice of $n$, and $g_{BL}$ define a theory which is bounded from above. However, regardless of the value of the $B-L$ charge, and the choice of $g_{BL}$, we note that there is an absolute upper bound at the multi-TeV scale for the seesaw scale. This statement may not be trivially seen in Fig.~6 and 7, because it seems that the larger the $B-L$ charge, and the coupling $g_{BL}$, the larger the upper bound on the $B-L$ breaking scale. However, the coupling $g_{BL}$ is bounded by perturbativity, and in the limit of large $n$, the annihilation into two fermions, which defines the upper bound on the resonance, becomes insensitive to the $B-L$ charge for large values thereof. We note that the annihilation channel into to new gauge bosons is irrelevant for defining the upper bound since it is bounded by perturbativity of the $B-L$ coupling. These results are crucial to understand the testability of this theory at colliders. Clearly, we could test at the Large Hadron Collider only one fraction of the parameter space. 

For completeness of the discussion on the upper bound for the seesaw scale, we would like to mention that the upper bound coming from the cosmological bound on the relic density is in agreement with partial-wave unitarity of the S-matrix. It is well known that, from a naive model-independent study, partial wave unitarity requires that $M_\chi < 340$ TeV~\cite{Griest:1989wd}. However, in this model, the partial wave expansion only becomes relevant in regions of the parameter space which are not allowed by cosmology. Therefore, unitarity of the S-matrix does not make any influence on the upper bound for the seesaw scale.

The main implication of these results is that there is a hope to test the existence of lepton number violation since the upper bound on the $B-L$ breaking scale is much smaller than the canonical seesaw scale.

\begin{figure}[t]
\includegraphics[width=0.8\linewidth]{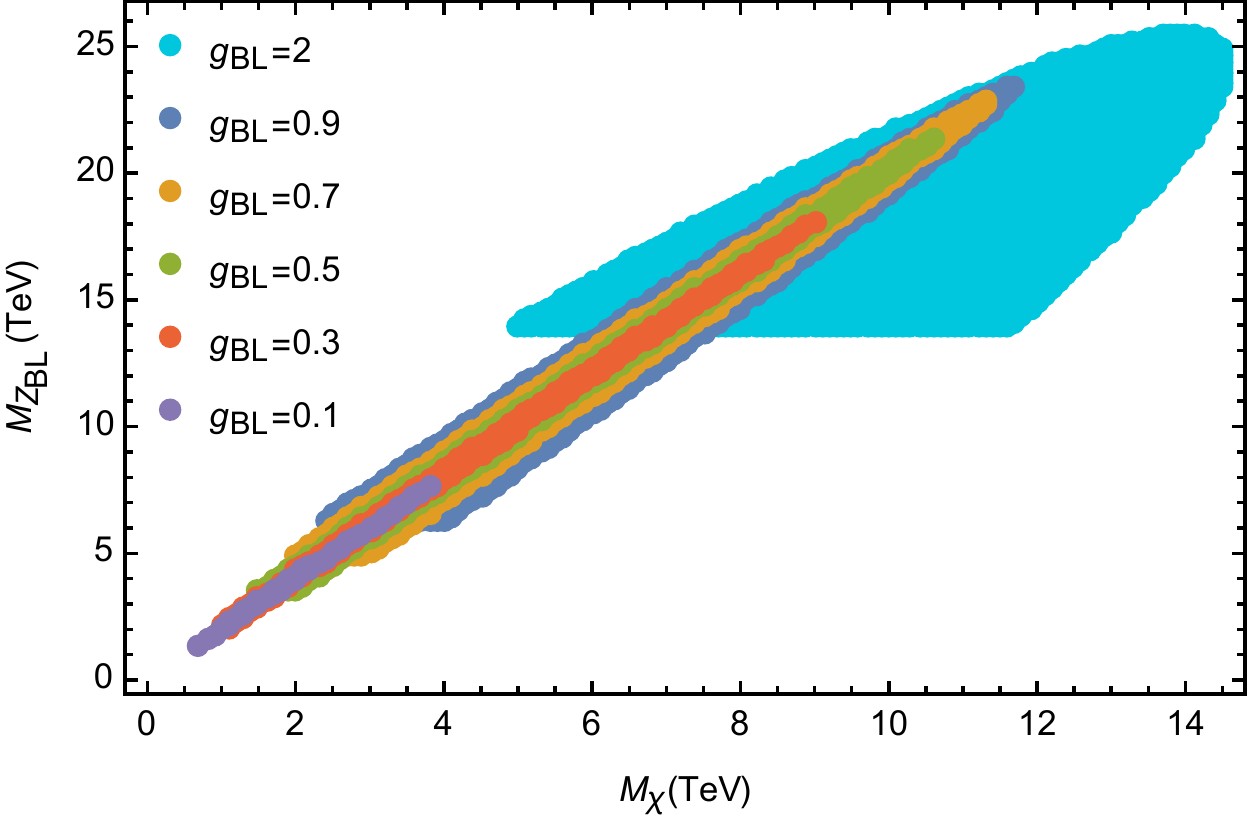}
\caption{Allowed region in the $M_{Z_{BL}}-M_\chi$ plane for $n=1/3$ when $\Omega_{DM} h^2 \leq 0.1199 \pm 0.0027$, in agreement with the LEP bounds. Here we use the perturbative bound $g_{BL} < 2 \sqrt{\pi}$.}
\label{RD_allg}
\end{figure}

\begin{figure}[t]
\includegraphics[width=0.8\linewidth]{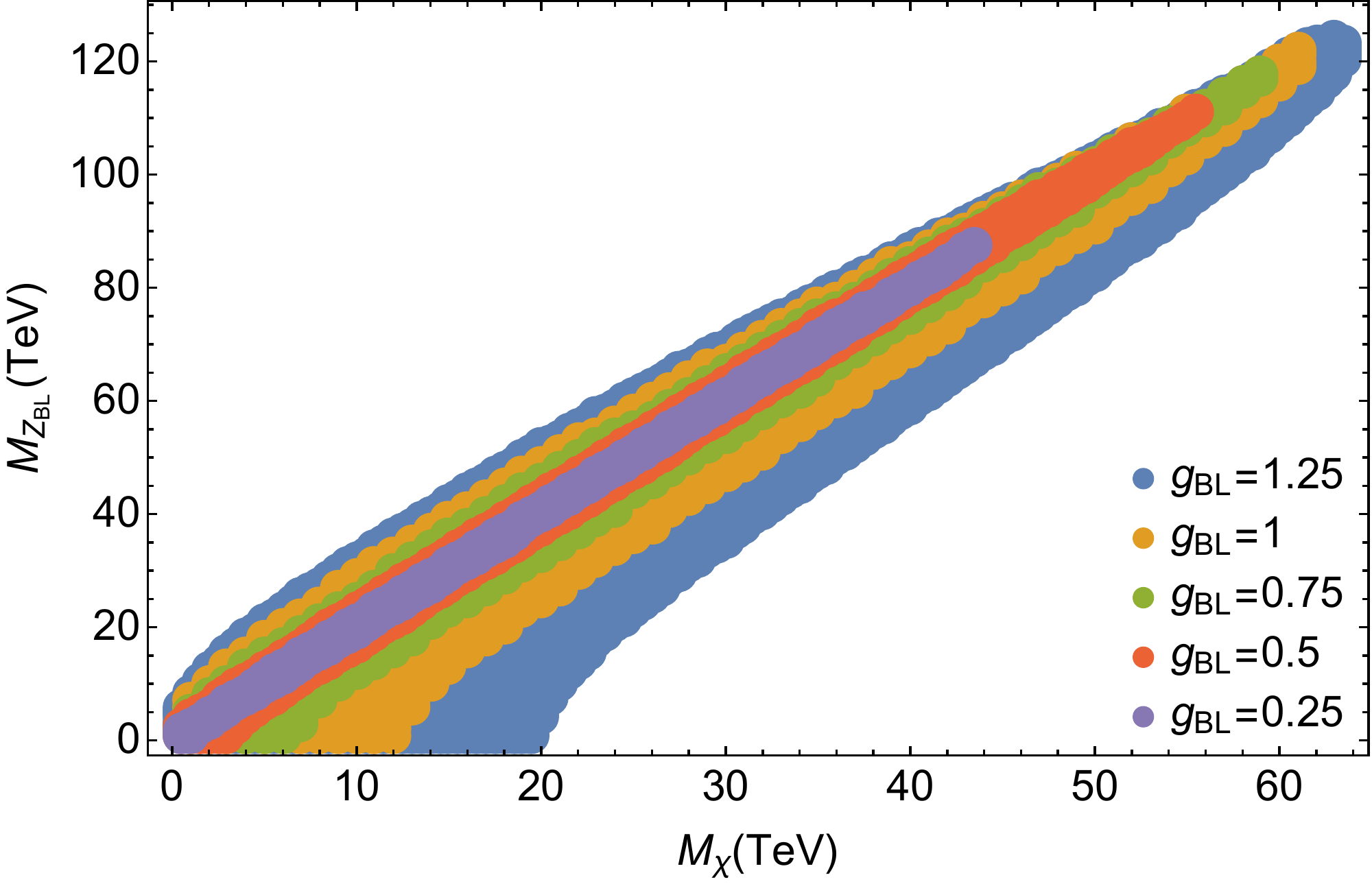}
\caption{Allowed region in the $M_{Z_{BL}}-M_\chi$ plane for $n=2$ when $\Omega_{DM} h^2 \leq 0.1199 \pm 0.0027$, in agreement with the LEP bounds. Here we use the perturbative bound $g_{BL} < \sqrt{\pi}$.}
\label{RD_allg}
\end{figure}

\end{itemize}

\section{Lepton Number Violation at the LHC}
\begin{figure}[t]
\includegraphics[width=0.8\linewidth]{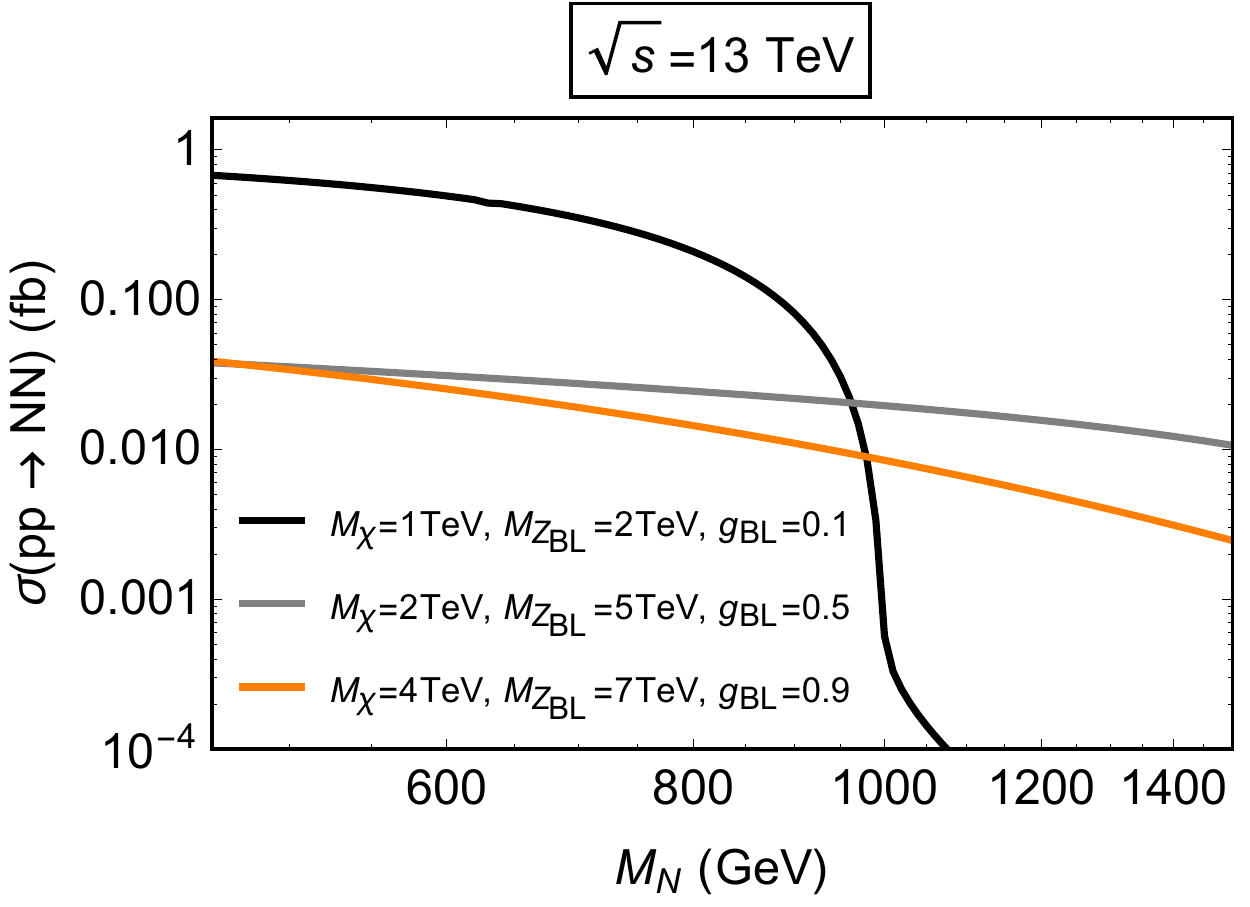}
\includegraphics[width=0.8\linewidth]{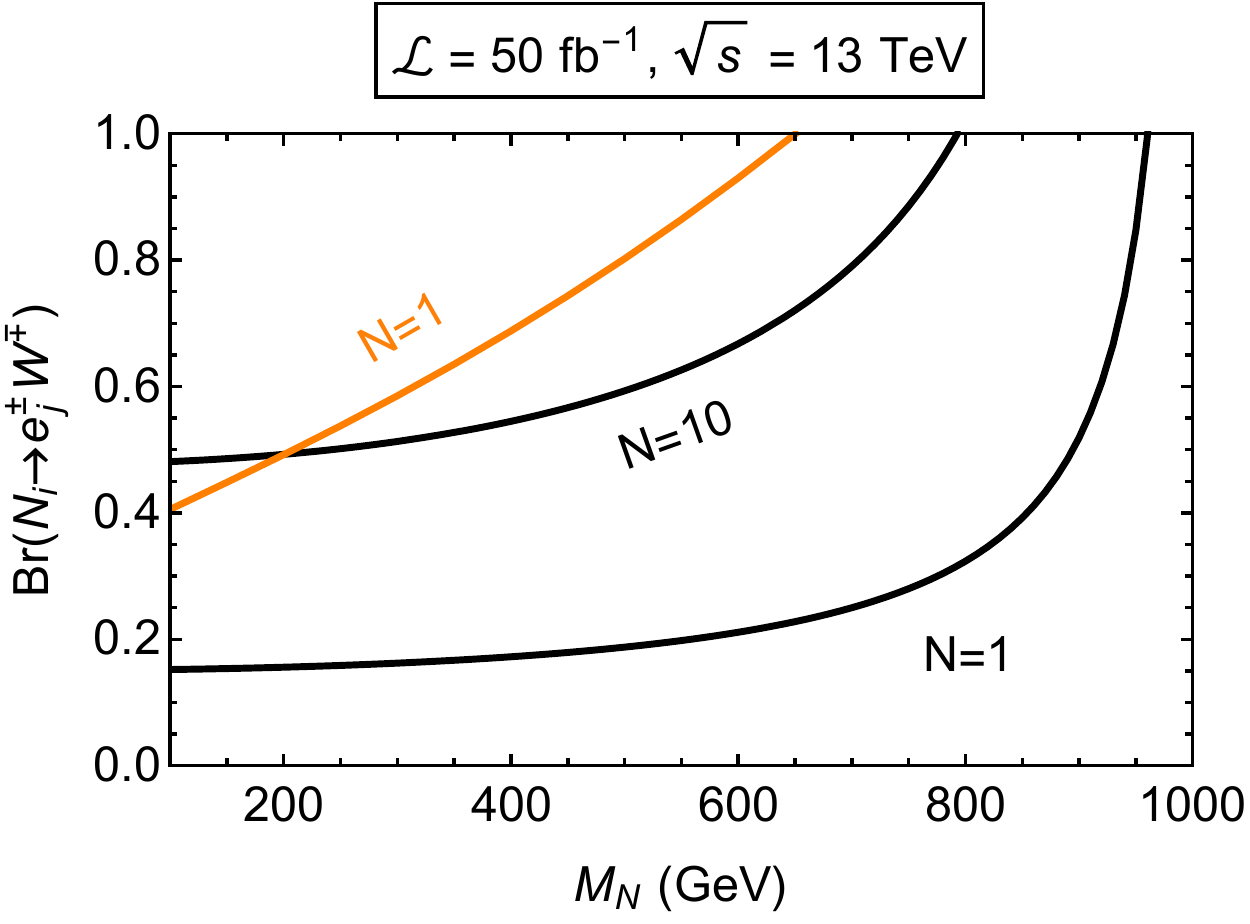}
\caption{In upper-panel, production cross section for the right-handed neutrinos at the LHC when $\sqrt{s}=13$ TeV, and in different scenarios consistent with the dark matter relic density. In the lower-panel, expected number of events for the scenarios $M_\chi=1$ TeV, $M_{Z_{BL}}=2$ TeV, $g_{BL}=0.1$ (black), and $M_\chi=4$ TeV, $M_{Z_{BL}}=7$ TeV, $g_{BL}=0.9$ (orange), when $\sqrt{s}=13$ TeV, and ${\cal L}=50 \ \text{fb}^{-1}$.}
\label{comparison}
\end{figure}
In the previous section we have shown the possibility to find an upper bound on the $B-L$ symmetry breaking scale using the constraints from the dark matter relic density. Therefore, one can hope to test the existence of a new force associated to $B-L$, and observe lepton number violation at the LHC. Unfortunately, the upper bound is large if $n=2$, but still we can hope to test this theory if the symmetry is broken much below the upper bound. 

The right-handed neutrinos can be produced at the LHC through the neutral gauge boson $Z_{BL}$, i.e. $pp \to Z_{BL}^* \to N_i N_i$ or through the Higgses present in the theory
~\cite{Basso:2008iv,Huitu:2008gf,AguilarSaavedra:2009ik,Accomando:2017qcs,Perez:2009mu}. Since the production mechanisms through the Higgses suffer from the dependence on the mixing angle in the Higgs sector, we focus our discussion on the production through the $B-L$ gauge boson. 
The right-handed neutrinos could have the following decays: $$N_i \to e^{\pm}_j W^{\mp}, \nu_j Z, \nu_j h_1, \nu_j h_2.$$ Therefore, the lepton number violating signatures can be observed when the right-handed neutrinos decay into charged leptons, and one has the following channels with the same-sign leptons
\begin{equation}
p p \to Z_{BL}^* \to N_i N_i \to e^{\pm}_j W^{\mp} e^{\pm}_k W^{\mp} \to e^{\pm}_j e^{\pm}_k 4 j.
\end{equation}
The number of these events is given by
\begin{eqnarray}
N_{e^{\pm}_j e^{\pm}_k 4j} &=& 2 \times {\cal{L}} \times \sigma (pp \to N_i N_i) \times {\rm{Br}} (N_i \to e^{\pm}_j W^{\mp} ) \nonumber \\
&\times & {\rm{Br}} (N_i \to e^{\pm}_k W^{\mp} ) \times {\rm{Br}} (W \to jj)^2,
\end{eqnarray}
where ${\cal{L}}$ is the integrated luminosity. Here, the factor two is included to discuss the channels with same-sign leptons without distinguishing the electric charge of the leptons in the final state.

In Fig.~8 (upper-panel), we show the predictions for the cross section when $\sqrt{s}=13$ TeV, and for different scenarios consistent with relic density bounds. 
Here we choose $n=1/3$, and numerical values for $M_\chi$, $M_{Z_{BL}}$, and $g_{BL}$ which satisfy the cosmological bounds. In the lower-panel, we show the 
number of events assuming a luminosity ${\cal{L}}=50 \ \rm{fb}^{-1}$. We have reviewed the current LHC bounds and unfortunately they cannot exclude the 
region of the parameter space studied here. A more detailed experimental study will help to understand the testability of this theory with more luminosity. 
Similar results are shown in Fig.~9, when $\sqrt{s}=14$ TeV. However, in this case one expects a large number of events for the same-sign leptons for right-handed neutrino masses below 1 TeV. 
\begin{figure}[t]
\includegraphics[width=0.8\linewidth]{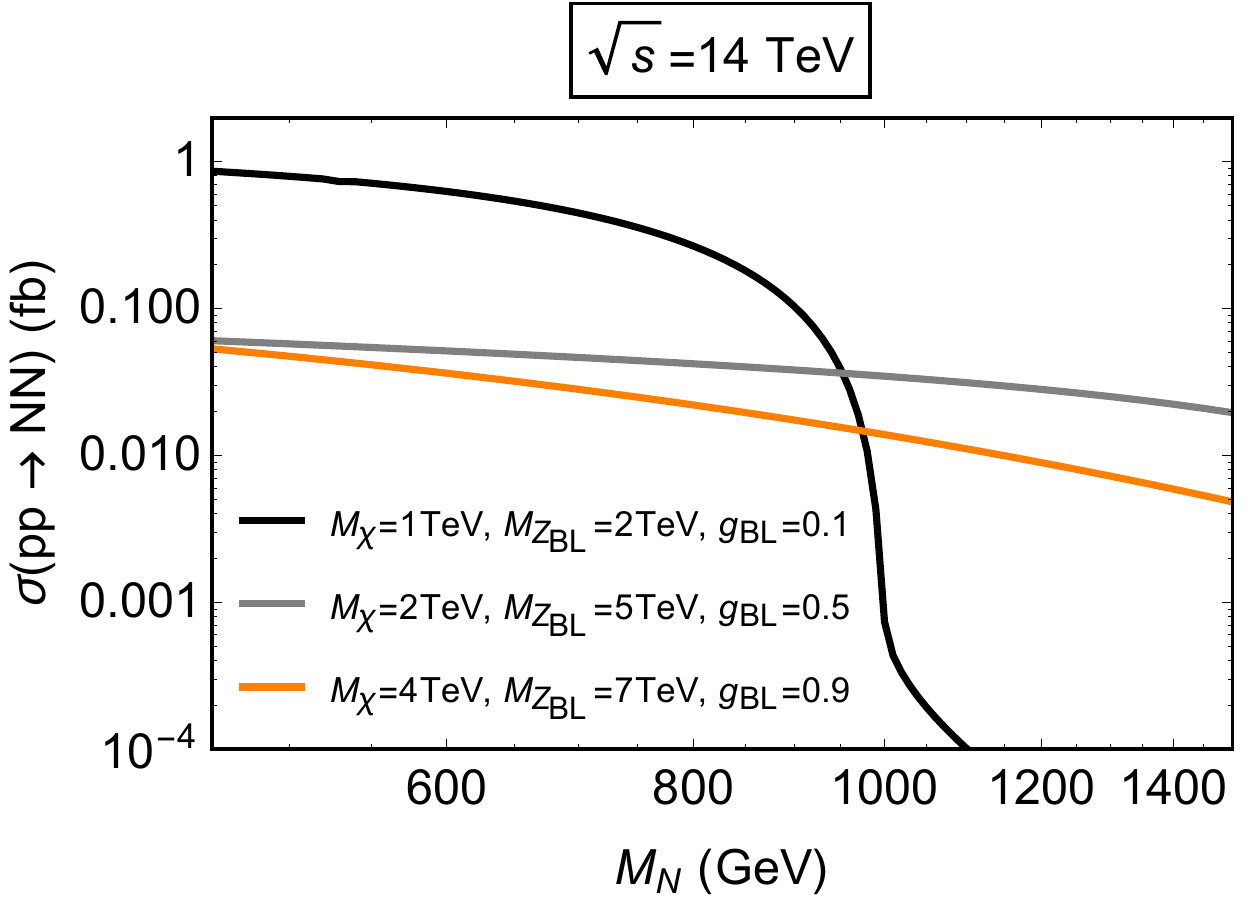}
\includegraphics[width=0.8\linewidth]{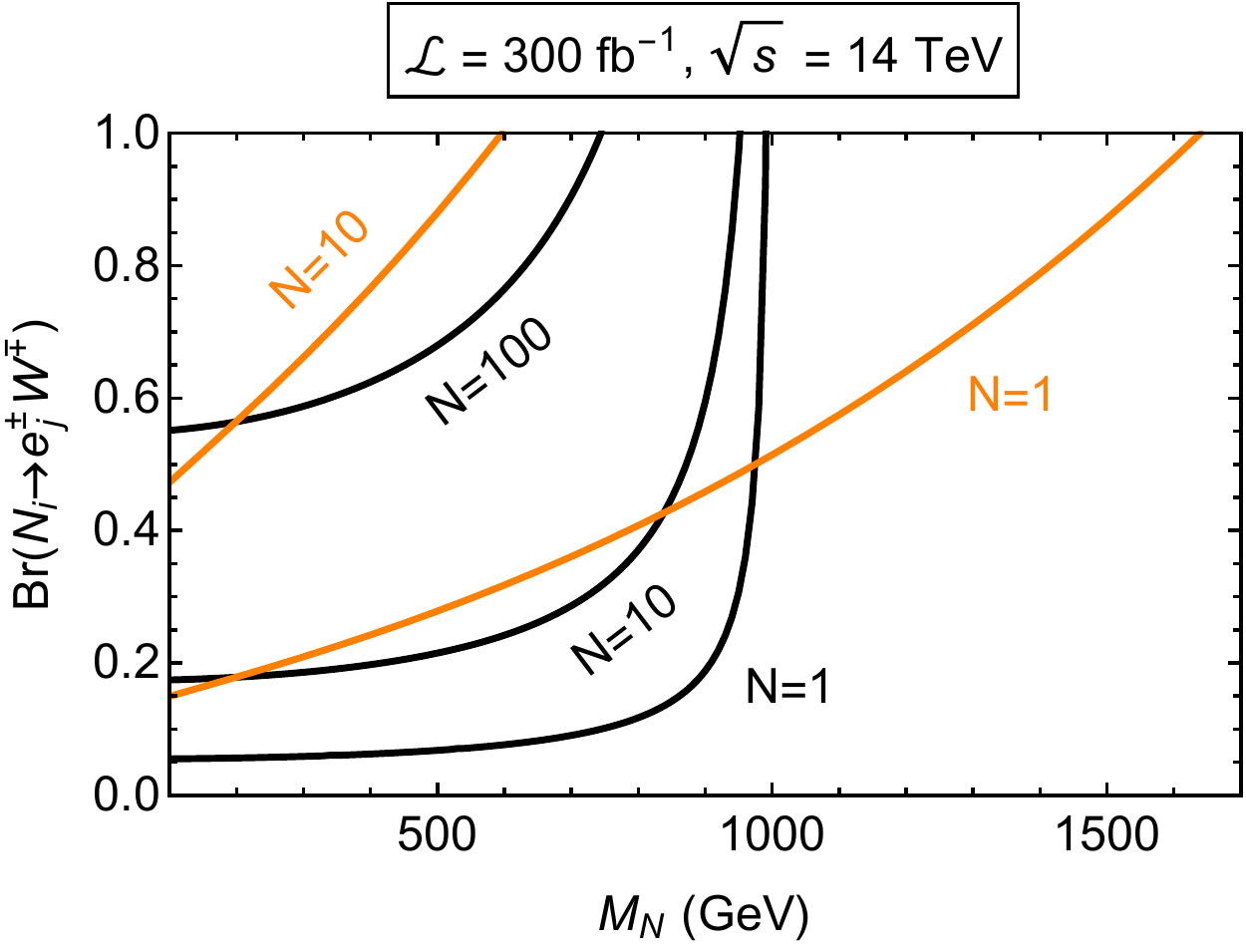}
\caption{In the upper-panel, production cross section for the right-handed neutrinos at the LHC when $\sqrt{s}=14$ TeV, and in different scenarios consistent with the dark matter relic density. In the lower-panel, expected number of events for the scenarios $M_\chi=1$ TeV, $M_{Z_{BL}}=2$ TeV, $g_{BL}=0.1$ (black), and $M_\chi=4$ TeV, $M_{Z_{BL}}=7$ TeV, $g_{BL}=0.9$ (orange), when $\sqrt{s}=14$ TeV, and ${\cal L}=300 \ \text{fb}^{-1}$.}
\label{comparison}
\end{figure}

In Fig.~10, we show the predictions for the next generation of proton-proton collider at 100 TeV. 
In this case one should be able to test most of the parameter space even when the right-handed neutrinos are in the multi-TeV region.
See Ref.~\cite{Arkani-Hamed:2015vfh} for the discovery potential of the 100 TeV collider. We would like to mention that, in a large part of the parameter space, the right handed neutrino decays can give rise to displaced vertices~\cite{Perez:2009mu}. Since the seesaw scale has to be at most at the multi-TeV scale $\sim {\cal O}(10^2$ TeV), the Yukawa Dirac $Y_\nu$ has to be small in order to reproduce the measured active neutrino masses. A small $Y_\nu$ enhances the lifetime of the right-handed neutrinos, which then become long-lived particles. Therefore, as a consequence of having a low seesaw scale, displaced vertices arise as an exotic signatures predicted by the model. 
For instance, when the right-handed neutrino mass is about $400$ GeV, the decay length can be 
\begin{displaymath}
L=(10^{-3}-10^{-1}) \rm{mm}.
\end{displaymath} 
This simple study motivates the search of lepton number violating signatures at the LHC or at future colliders.
\begin{figure}[t]
\includegraphics[width=0.8\linewidth]{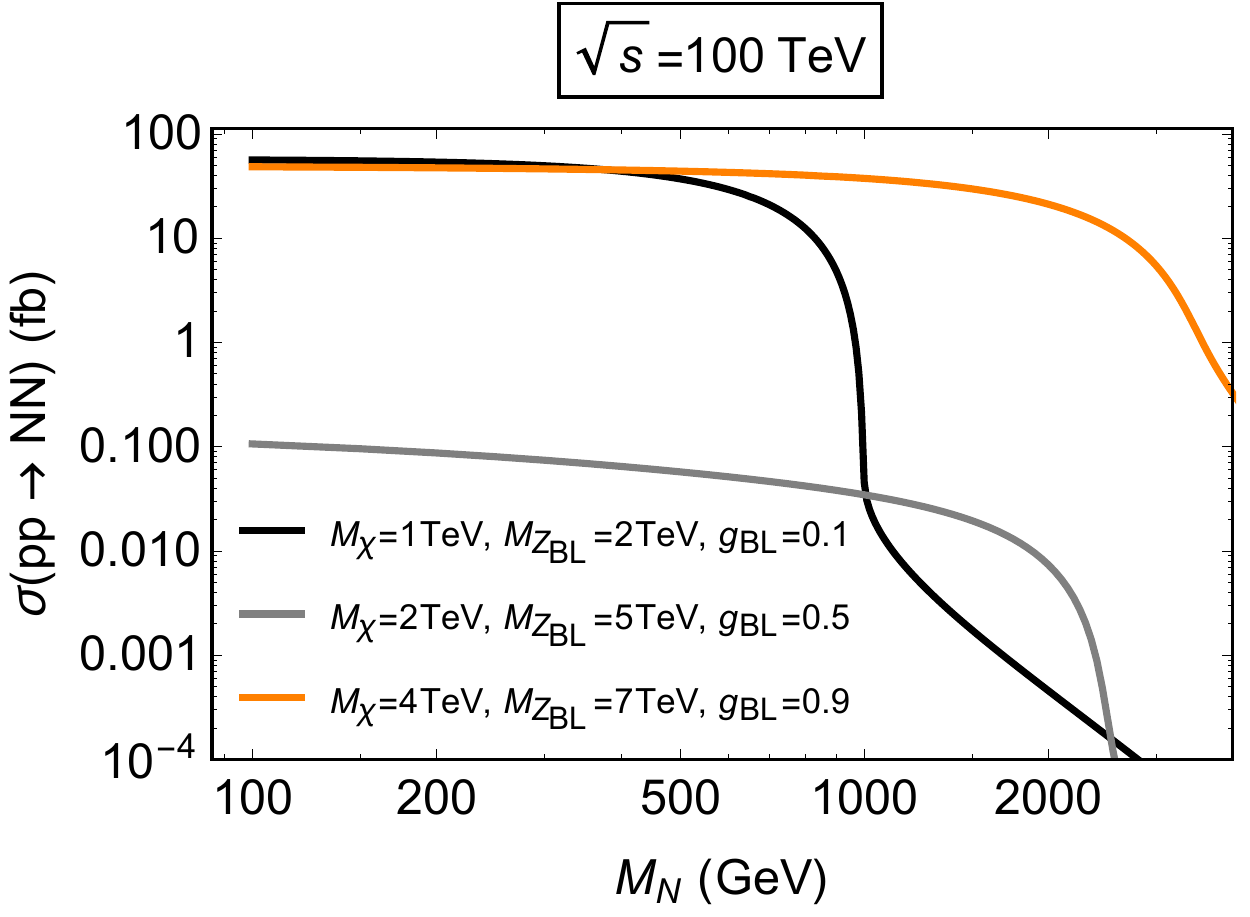}
\includegraphics[width=0.8\linewidth]{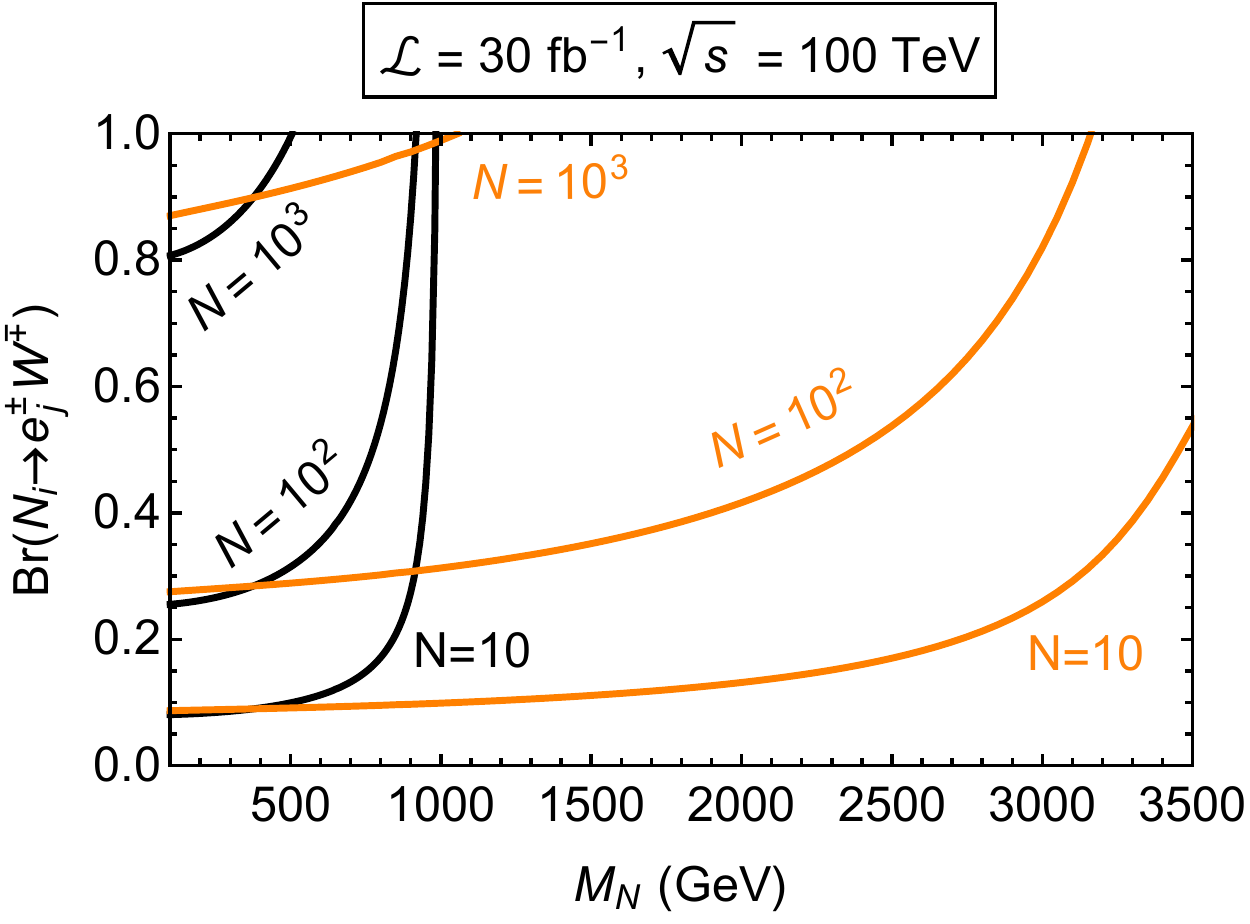}
\caption{Production cross section for the right-handed neutrinos at the 100 TeV collider and in different scenarios consistent with the dark matter relic density (left panel), and the expected number of events for the scenarios $M_\chi=1$ TeV, $M_{Z_{BL}}=2$ TeV, $g_{BL}=0.1$ (black) and $M_\chi=4$ TeV, $M_{Z_{BL}}=7$ TeV, $g_{BL}=0.9$ (orange), when $\sqrt{s}=100$ TeV, and ${\cal L}=30 \ \text{fb}^{-1}$ (right-panel).}
\label{comparison}
\end{figure}
\section{Final Discussion}
We have discussed  the simple seesaw mechanism for neutrino masses where the seesaw scale is defined by the $B-L$ symmetry breaking scale and 
the scenarios where it is possible to find an upper bound on the seesaw scale using theoretical arguments or cosmological bounds.
We have investigated the relation between the origin of neutrino masses and the properties of a simple cold dark matter candidate in the 
context of a theory based on the $B-L$ gauge symmetry. In this context the upper bound on the seesaw scale is in the multi-TeV region and one predicts the existence of exotic lepton number violating signatures at colliders. We use all cosmological constraints and 
investigate the predictions for direct and indirect detection dark matter experiments. These cosmological bounds motivate the search for lepton number violation at the Large Hadron Collider or at future colliders in order to test the origin of neutrino masses.
\\
\\
{\textit{Acknowledgments:}} P.~F.~P.  thanks Mark B. Wise for discussions and the Walter Burke Institute for Theoretical Physics at Caltech for hospitality. C.~M. thanks the theory division at IFIC for fruitful comments and discussions. The work of P.~F.~P. has been supported by the U.S. Department of Energy under contract No. de-sc0018005. The work of C.~M. has been supported in part by the Spanish Government and ERDF funds from the EU Commission [Grants No. FPA2014-53631-C2-1-P and SEV-2014- 0398] and "La Caixa-Severo Ochoa" scholarship.
\newpage
\appendix
\begin{widetext}
\section{Appendix A. Dark Matter Annihilation Cross Sections}
The annihilation cross section for $\bar{\chi} \chi \ \to \ Z_{BL}^\ast \  \to \ \bar{f} f$ is given by
\begin{equation}
\label{eq:AnnihilationCrossSection}
\sigma (\bar{\chi} \chi \to  Z_{BL}^\ast   \to  \bar{f} f) = \frac{N_c^{f} n_f^2 g_{BL}^4 n^2 }{12 \pi s} \frac{\sqrt{s - 4 M_{f}^2}}{\sqrt{s - 4 M_\chi^2}} \frac{\left( s + 2 M_\chi^2\right) \left( s + 2 M_{f}^2 \right)}{ \left[ (s- M_{Z_{BL}}^2)^2 + M_{Z_{BL}}^2 \Gamma_{Z_{BL}}^2 \right]}.
\end{equation}
Here $N_c^f$ is the color factor of the fermion $f$ with mass $M_f$, $s$ is the square of the center-of-mass energy, and $\Gamma_{Z_{BL}}$ is the total decay width of the $Z_{BL}$ gauge boson. For Majorana neutral fermions, the annihilation cross section into fermions reads as 
\begin{equation}
\label{eq:AnnihilationCrossSection}
\sigma (\bar{\chi} \chi \to  Z_{BL}^\ast   \to  \bar{f} f) = \frac{N_c^{f} n_f^2 g_{BL}^4 n^2 }{24 \pi s} \frac{\sqrt{s - 4 M_{f}^2}}{\sqrt{s - 4 M_\chi^2}} \frac{\left( s + 2 M_\chi^2\right) \left( s - 4 M_{f}^2 \right)}{ \left[ (s- M_{Z_{BL}}^2)^2 + M_{Z_{BL}}^2 \Gamma_{Z_{BL}}^2 \right]}.
\end{equation}

The annihilation cross section for  $\bar{\chi} \chi \ \to \ Z_{BL} Z_{BL}$ is given by
\begin{equation}
\begin{split}
&\sigma (\bar{\chi} \chi \to Z_{BL} Z_{BL})=\frac{g_{BL}^4n^{4}}{32 \pi E^2} \frac{\omega}{v}\left[ -1 -\frac{(2+z^2)^2}{(2-z^2)^2-4 v^2}+\frac{6-2z^2+z^4+12v^2+4v^4}{2v\omega(1+v^2+ \omega^2)}\ln \left(\frac{1+(v+\omega)^2}{1+(v-\omega)^2}\right)\right],
\end{split}
\end{equation}
where $z$, $v$ and $\omega$ are defined as
\begin{eqnarray}
z&=&M_{Z_{BL}}/M_\chi, \ 
v= p/M_\chi,\  \text{and} \ 
\omega = k/M_\chi,
\end{eqnarray}
being $E$ and $p$ the center-of-mass energy and momentum of the initial particles respectively, $E=\sqrt{s}/2$ and $p=\sqrt{E^2-M_\chi^2}$, and $k$ the momentum of the final particles, $k=\sqrt{E^2-M_{Z_{BL}}^2}$. The annihilation cross section $\bar{\chi} \chi \to Z_{BL} h_{i}$ is given by
\begin{equation}
\begin{split}
\sigma(\bar{\chi} \chi \to Z_{BL} h_{i})&= c_i^2 \frac{g_{BL}^4 n_S^2 n^2}{48 \pi s^2}\left(1+\displaystyle \frac{2M_\chi^2}{s}\right )\frac{(s^2+2s(5M_{Z_{BL}}^2-M_{h_i}^2)+(M_{Z_{BL}}^2-M_{h_i}^2)^2)}{(s-M_{Z_{BL}}^2)^2+\Gamma_{Z_{BL}}^2M^2_{Z_{BL}}}\\
& \times \frac{\sqrt{(s+M_{Z_{BL}}^2-M_{h_{i}}^2)^2-4M_{Z_{BL}}^2 s}}{\sqrt{1-\displaystyle \frac{4M_\chi^2}{s}}},
\end{split}
\end{equation}
where $c_i=(\sin \theta_{BL}, \cos \theta_{BL})$ and $n_S$ is the $B-L$ charge of the $S_{BL}$ Higgs.
%
\section{Appendix B. $Z_{BL}$ Decays}
For the decays into charged fermions:
\begin{equation}
\Gamma(Z_{BL}\to \bar{f}_i f_i)=\frac{g_{BL}^2N_c n_{f_i}^2}{12\pi M_{Z_{BL}}}\sqrt{1-\frac{4 M_{f_i}^2}{M_{Z_{BL}}^2}}(2 M_{f_i}^2+M_{Z_{BL}}^2).
\end{equation}
For Majorana neutral fermions:
\begin{eqnarray}
\Gamma(Z_{BL}\to \nu \nu)&=&\frac{g_{BL}^2 }{24\pi}M_{Z_{BL}},\\
\Gamma(Z_{BL}\to NN)&=&\frac{g_{BL}^2}{24\pi M_{Z_{BL}}}\sqrt{1-\frac{4 M_N^2}{M_{Z_{BL}}^2}}(M_{Z_{BL}}^2-4 M_N^2).
\end{eqnarray}
Total decay width of the $Z_{BL}$:
\begin{equation}
\begin{split}
\Gamma_\text{tot}(Z_{BL})&=3 \, \Gamma(Z_{BL}\to \ell \ell) + 5 \, \Gamma(Z_{BL}\to \bar{q} q) + \Gamma(Z_{BL}\to \bar{t} t)\\
&+3 \Gamma(Z_{BL}\to  \nu \nu) +3   \Gamma(Z_{BL}\to NN) +\Gamma(Z_{BL}\to \bar{\chi} \chi).
\end{split}
\end{equation}
\section{Appendix C. Right-handed Neutrinos Production Cross Section}
%
The parton level cross section for this process is
\begin{equation}
{d\sigma(q\bar{q}\to Z_{BL}^* \to N_i N_i)\over dt}={1\over 32\pi
s(s-4M_q^2)N_c}{2g_{BL}^4\over 9} \frac{\left[ \left(s+t-2 M_{N_i}^2\right)^2+(t-2M_q^2)^2-2( M_{N_i}^2+M_q^2)^2 \right]}{ (s-M_{Z_{BL}}^2)^2+M_{Z_{BL}}^2 \Gamma^2_{Z_{BL}}}, \nonumber 
\end{equation}
where $t=(p_q-p_N)^2$ and the color factor for quarks is $N_c=3$. The integrated expression reads as
\begin{equation}
\sigma(q\bar{q}\to Z_{BL}^* \to N_i N_i)=\frac{g_{BL}^4(s-4 M_{N_i}^2)(s+2M_q^2)}{648\pi s\left((s-M_{Z_{BL}}^2)^2+\Gamma_{Z_{BL}}^2 M_{Z_{BL}}^2\right)}\sqrt{\frac{s-4 M_{N_i}^2}{s-4M_q^2}}.
\end{equation}

\end{widetext}

%



\begin{thebibliography}{99}
%

\bibitem{FileviezPerez:2009ym}
  P.~Fileviez Perez,
  ``The Origin of Neutrino Masses and Physics Beyond the Standard Model,''
  AIP Conf.\ Proc.\  {\bf 1222} (2010) 3
  doi:10.1063/1.3399352
  [arXiv:0909.2698 [hep-ph]].

 \bibitem{TypeI}
  P.~Minkowski,
  ``$\mu \to e$ Gamma At A Rate Of One Out Of 1-Billion Muon Decays?,''
  Phys.\ Lett.\ B {\bf 67} (1977) 421;
  R.~N.~Mohapatra and G.~Senjanovic,
  ``Neutrino Mass and Spontaneous Parity Violation,''
  Phys.\ Rev.\ Lett.\  {\bf 44} (1980) 912;
  T. Yanagida, in {\it Proceedings of the Workshop on the Unified Theory
   and the Baryon Number in the Universe}, eds. O. Sawada et al., (KEK
   Report~79-18, Tsukuba, 1979), p.~95;
  M. Gell-Mann, P. Ramond and R. Slansky,
   in {\it Supergravity}, eds. P. van Nieuwenhuizen et al.,
   (North-Holland, 1979), p.~315;
  S.L. Glashow, in {\it Quarks and Leptons}, Carg\`ese, eds. M. L\'evy et al.,
(Plenum, 1980), p. 707. 


\bibitem{Bertone:2004pz}
  G.~Bertone, D.~Hooper and J.~Silk,
  ``Particle dark matter: Evidence, candidates and constraints,''
  Phys.\ Rept.\  {\bf 405} (2005) 279
  doi:10.1016/j.physrep.2004.08.031
  [hep-ph/0404175].
  
\bibitem{Elliott:2002xe}
  S.~R.~Elliott and P.~Vogel,
  ``Double beta decay,''
  Ann.\ Rev.\ Nucl.\ Part.\ Sci.\  {\bf 52} (2002) 115
  doi:10.1146/annurev.nucl.52.050102.090641
  [hep-ph/0202264].
  
\bibitem{Cai:2017mow}
  W.~Y.~Keung and G.~Senjanovic,
  ``Majorana Neutrinos and the Production of the Right-handed Charged Gauge Boson,''
  Phys.\ Rev.\ Lett.\  {\bf 50} (1983) 1427.
  Y.~Cai, T.~Han, T.~Li and R.~Ruiz,
  ``Lepton-Number Violation: Seesaw Models and Their Collider Tests,''
  arXiv:1711.02180 [hep-ph].
  
\bibitem{Barger:2008wn}
  V.~Barger, P.~Fileviez Perez and S.~Spinner,
  ``Minimal gauged U(1)(B-L) model with spontaneous R-parity violation,''
  Phys.\ Rev.\ Lett.\  {\bf 102} (2009) 181802
  doi:10.1103/PhysRevLett.102.181802
  [arXiv:0812.3661 [hep-ph]].
  
  
\bibitem{Basso:2008iv}
  L.~Basso, A.~Belyaev, S.~Moretti and C.~H.~Shepherd-Themistocleous,
  ``Phenomenology of the minimal B-L extension of the Standard model: Z' and neutrinos,''
  Phys.\ Rev.\ D {\bf 80} (2009) 055030
  doi:10.1103/PhysRevD.80.055030
  [arXiv:0812.4313 [hep-ph]].
  
\bibitem{Huitu:2008gf}
  K.~Huitu, S.~Khalil, H.~Okada and S.~K.~Rai,
  ``Signatures for right-handed neutrinos at the Large Hadron Collider,''
  Phys.\ Rev.\ Lett.\  {\bf 101} (2008) 181802
  doi:10.1103/PhysRevLett.101.181802
  [arXiv:0803.2799 [hep-ph]].
  

\bibitem{AguilarSaavedra:2009ik}
  J.~A.~Aguilar-Saavedra,
  ``Heavy lepton pair production at LHC: Model discrimination with multi-lepton signals,''
  Nucl.\ Phys.\ B {\bf 828} (2010) 289
  doi:10.1016/j.nuclphysb.2009.11.021
  [arXiv:0905.2221 [hep-ph]].
  
\bibitem{Accomando:2017qcs}
  E.~Accomando, L.~Delle Rose, S.~Moretti, E.~Olaiya and C.~H.~Shepherd-Themistocleous,
  ``Extra Higgs Boson and $Z'$ as Portals to Signatures of Heavy Neutrinos at the LHC,''
  arXiv:1708.03650 [hep-ph].
    
\bibitem{Perez:2009mu}
  P.~Fileviez Perez, T.~Han and T.~Li,
  ``Testability of Type I Seesaw at the CERN LHC: Revealing the Existence of the B-L Symmetry,''
  Phys.\ Rev.\ D {\bf 80} (2009) 073015
  doi:10.1103/PhysRevD.80.073015
  [arXiv:0907.4186 [hep-ph]].



\bibitem{FileviezPerez:2012mj}
  P.~Fileviez Perez and S.~Spinner,
  ``The Minimal Theory for R-parity Violation at the LHC,''
  JHEP {\bf 1204} (2012) 118
  doi:10.1007/JHEP04(2012)118
  [arXiv:1201.5923 [hep-ph]].

\bibitem{FileviezPerez:2008sx}
  P.~Fileviez Perez and S.~Spinner,
  ``Spontaneous R-Parity Breaking and Left-Right Symmetry,''
  Phys.\ Lett.\ B {\bf 673} (2009) 251
  doi:10.1016/j.physletb.2009.02.047
  [arXiv:0811.3424 [hep-ph]].

\bibitem{Perez:2013kla}
  P.~Fileviez Perez and S.~Spinner,
  ``Supersymmetry at the LHC and The Theory of R-parity,''
  Phys.\ Lett.\ B {\bf 728} (2014) 489
  doi:10.1016/j.physletb.2013.12.022
  [arXiv:1308.0524 [hep-ph]].

\bibitem{Barger:2010iv}
  V.~Barger, P.~Fileviez Perez and S.~Spinner,
  ``Three Layers of Neutrinos,''
  Phys.\ Lett.\ B {\bf 696} (2011) 509
  doi:10.1016/j.physletb.2011.01.015
  [arXiv:1010.4023 [hep-ph]].
 
 
\bibitem{Duerr:2015wfa}
  M.~Duerr, P.~Fileviez Perez and J.~Smirnov,
  ``Simplified Dirac Dark Matter Models and Gamma-Ray Lines,''
  Phys.\ Rev.\ D {\bf 92} (2015) no.8,  083521
  doi:10.1103/PhysRevD.92.083521
  [arXiv:1506.05107 [hep-ph]].
 
\bibitem{Gondolo:1990dk}
  P.~Gondolo and G.~Gelmini,
  ``Cosmic abundances of stable particles: Improved analysis,''
  Nucl.\ Phys.\ B {\bf 360} (1991) 145.
 
\bibitem{RelicDensity}
  P.~A.~R.~Ade {\it et al.} [Planck Collaboration],
  ``Planck 2013 results. XVI. Cosmological parameters,''
  Astron.\ Astrophys.\  {\bf 571} (2014) A16
  doi:10.1051/0004-6361/201321591
  [arXiv:1303.5076 [astro-ph.CO]].
  
\bibitem{Alioli:2017nzr}
  S.~Alioli, M.~Farina, D.~Pappadopulo and J.~T.~Ruderman,
  ``Catching a New Force by the Tail,''
  arXiv:1712.02347 [hep-ph].
  

\bibitem{Aprile:2018dbl}
  E.~Aprile {\it et al.} [XENON Collaboration],
  ``Dark Matter Search Results from a One Tonne$\times$Year Exposure of XENON1T,''
  arXiv:1805.12562 [astro-ph.CO].
  
\bibitem{Aprile:2017iyp}
  E.~Aprile {\it et al.} [XENON Collaboration],
  ``First Dark Matter Search Results from the XENON1T Experiment,''
  Phys.\ Rev.\ Lett.\  {\bf 119} (2017) no.18,  181301
  doi:10.1103/PhysRevLett.119.181301
  [arXiv:1705.06655 [astro-ph.CO]].
  
\bibitem{Aprile:2015uzo}
  E.~Aprile {\it et al.} [XENON Collaboration],
  ``Physics reach of the XENON1T dark matter experiment,''
  JCAP {\bf 1604} (2016) no.04,  027
  doi:10.1088/1475-7516/2016/04/027
  [arXiv:1512.07501 [physics.ins-det]].
  
  
  
\bibitem{Ackermann:2015zua}
  M.~Ackermann {\it et al.} [Fermi-LAT Collaboration],
  ``Searching for Dark Matter Annihilation from Milky Way Dwarf Spheroidal Galaxies with Six Years of Fermi Large Area Telescope Data,''
  Phys.\ Rev.\ Lett.\  {\bf 115} (2015) no.23,  231301
  doi:10.1103/PhysRevLett.115.231301
  [arXiv:1503.02641 [astro-ph.HE]].
 
\bibitem{Griest:1989wd}
  K.~Griest and M.~Kamionkowski,
  ``Unitarity Limits on the Mass and Radius of Dark Matter Particles,''
  Phys.\ Rev.\ Lett.\  {\bf 64} (1990) 615.
  doi:10.1103/PhysRevLett.64.615
  
\bibitem{Arkani-Hamed:2015vfh}
  N.~Arkani-Hamed, T.~Han, M.~Mangano and L.~T.~Wang,
  ``Physics opportunities of a 100 TeV proton?proton collider,''
  Phys.\ Rept.\  {\bf 652} (2016) 1
  doi:10.1016/j.physrep.2016.07.004
  [arXiv:1511.06495 [hep-ph]].
 
  
\end{thebibliography}
\end{document}